# Test of transient deviations from Quantum Mechanics in Bell's experiment.

Alejandro Hnilo, Mónica Agüero, Marcelo Kovalsky and Myriam Nonaka.

*CEILAP, Centro de Investigaciones en Láseres y Aplicaciones, UNIDEF (MINDEF-CONICET);*
*CITEDEF, J.B. de La Salle 4397, (1603) Villa Martelli, Argentina.*
*email: ahnilo@citedef.gob.ar*

April 15th, 2024.

The conflict between Quantum Mechanics (QM) and Local Realism is most noticeable in the correlations observed between distant regions of a spatially spread entangled state. It has been hypothesized that transient deviations (from the values predicted by QM) may be observed if the correlations are measured in a time shorter than $L/c$, where $L$ is the spatial spread of the entangled state and $c$ is the speed of light. This hypothesis is appealing for it solves that conflict by minimally modifying the interpretation of QM, and opens the door to potentially fruitful nonlinear generalizations of QM without the risk of allowing faster-than-light signaling. The hypothesis is technically impossible to test directly nowadays, but a stroboscopic test is attainable. We present the results of such a test performed on a specially designed optical Bell setup with a distance between stations up to 24 m in straight line. No difference with the same observations performed at short distance, or evidence of transient deviations, is found. Yet, several hypotheses are involved in this experiment; they are detailed and briefly discussed. To say the least, the space left for the hypothesis of transient deviations is much reduced.

Scientific knowledge is tentative and provisional. Some day in the future, experiments will show that Quantum Mechanics as we know it (QM) is an approximation to a more general (and probably even stranger) theory. Finding those experiments, i.e., finding the limits of validity of QM, is an important problem in Physics.

One of the proposed limits of validity of QM is the existence of *transient* deviations from its predictions [1]. QM predictions are statistical and have been always validated recording statistical averages, which require a "long" time to be collected. Records over "short" times might then show a deviation, other than mere statistical fluctuations, with QM predictions. Of course the meaning of "long" and "short" must be defined. As Locality is part of the discussions on the interpretations of QM, it is natural assuming the timescale to be the time $\tau \equiv L/c$ light needs to cover the spatial spread $L$ of the quantum state involved. In other words: QM would predict correct results if observations were recorded in a time $t/\tau \gg 1$, while deviations from these predictions would be observed in a time shorter than $\tau$. Note that, even if this transient deviations hypothesis (TDH) were true, QM predictions would be correct in the overwhelming number of cases.

An accessible quantum state to test TDH is a pair of photons entangled in polarization. Photons propagate at $c$, the value of $L$ can be made large, and observables which value is distinctive of "quantumness" are well known, f.ex., the $S_{CHSH}$ parameter [2]. At this point in the reasoning, the deviations may take values both smaller and higher than the QM prediction $S_{CHSH} = 2\sqrt{2}$. We make now the additional assumption that for $t \leq \tau$ the classical bound $S_{CHSH} \leq 2$ holds. This assumption is appealing for at least two reasons: Firstly, it would solve in a simple way the controversy between QM and Local Realism. Secondly, it would open the door to nonlinear extensions of QM [3] with no risk of allowing signals propagating faster than light [4,5].

The direct way to test the TDH is to observe the evolution of entanglement (say, $S_{CHSH}$) in a time shorter than $\tau$. But, this means a difficult challenge. For the fully symmetrical Bell's state, the maximum difference between the prediction of QM for the probability of coincidence, and the one of a classical theory, is $\approx 0.052$ (this occurs at the settings $\alpha-\beta= \pi/8$, $3\pi/8$). If the number of recorded coincidences is $N$, its statistical dispersion is $\approx \sqrt{N}$. Therefore, that difference is discernible only if $0.052 \times N \gg \sqrt{N}$, or $N \gg 368$. Therefore, roughly speaking, the number of coincidences should reach $\approx 10^3$ in each time resolved unit (or time slot) in order to be able to reliably observe that difference. The highest reported coincidence rate is $\approx 3\times 10^5$ s$^{-1}$ in a laboratory environment [6], what means $\approx 10^{-4}$ $\tau^{-1}$. Numbers are better for larger $L$: 50 s$^{-1}$ at 13 km [7] ($\approx 2\times 10^{-3}$ $\tau^{-1}$) and 8 s$^{-1}$ at 144 km [8] ($\approx 4\times 10^{-3}$ $\tau^{-1}$) but still short by several orders of magnitude, even if the value of the time slot is chosen equal to $\tau$ (the longest acceptable value).

An alternative at hand is to perform a stroboscopic observation, by using a pulsed source of entangled states. If the system decays to the same "ground state" before the arrival of the next pulse, and then follows the same evolution, then each pulse allows recording a single point (at random) of the hypothesized transient evolution. The complete evolution can be then reconstructed by collecting the points recorded in each time slot by as many pulses as necessary.

Numerical simulations of the consequences of the TDH, designed to be as general as possible, give an idea about how the transients may look like [9]. Transients may occur not only in $S_{CHSH}$, but also in the efficiency of detection $\eta$ (i.e., the rate of coincidences /singles). Depending on the parameters' values in the numerical simulation, monotonous evolution or oscillations with timescales longer than $\tau$ may occur in both $S_{CHSH}(t)$ and $\eta(t)$. The parameters' values are largely unknown, but in all cases, if $t/\tau \leq 1$:

$$S_{CHSH}(t) \times \eta(t) \leq 2 \qquad (1)$$

while $S_{CHSH}(t) \times \eta(t) \to 2\sqrt{2}$ for $t/\tau \gg 1$. This bound is recently demonstrated to be derivable from arithmetical properties of the time series of outcomes only, independently of the underlying model [10]. Eq.1 is then a general criterion to test the TDH.

A pertinent question is whether the TDH has been already disproved by the performed "loophole-free" experiments [11-17]. Those experiments are more sophisticated than the one presented here, but had a different purpose. The so-called loopholes are experimental deficiencies that allow some (rather conspiratorial) classical mechanisms *to fit* QM predictions. Instead, the TDH supposes QM predictions *to fail* during a short timescale. The loophole-free experiments measured time averaged observables in conditions such that the deficiencies were too small for the conspiratorial models to be able to succeed. Recording time behavior at the $\tau$ timescale was not among their aims. Instead, the experiment we describe in this paper is designed to reliably detect deviations at the $\tau$ timescale, and assumes the conspiratorial mechanisms to be nonexistent (see H3, H4 below). In fact, the results of the loophole-free experiments support this assumption.

Recently, we reported the stroboscopic observation of $S_{CHSH}(t)$ and $\eta(t)$ at different (long) $L$ values as measured through optical fibers [18], but at short distance in straight line. No transient deviations were visible. In this paper we report, for the first time, the results for the most relevant case: the evolution of $S_{CHSH}(t)$ and $\eta(t)$ with stations separated by a large distance as measured in straight line, and sufficient time resolution and statistics. Our setup has distinctive features, which are detailed in the Supplementary Material (SM) Section. In this main text we present and discuss only the most significant features and results.

A pulsed diode laser at 405 nm produces biphotons at 810 nm in the standard configuration with two crossed BBO-I crystals. The time values of photons' detections are recorded, and also the time of arrival of all laser pulses, even if no photon is detected during that pulse. These time series are recorded in two independent time-to-digital converters (TDCs), one in each station. They allow the (stroboscopic) reconstruction of the evolution of singles and coincidences. The position of each photon's detection inside the pulse is defined according to its time distance to the "trigger" signal provided by the nearest pulse's start.

The TDCs' nominal resolution is 10 ps, but the actual value is limited to $\approx 2$ ns because of jitter of single photon detectors. Time resolution defines the minimum value for $L/c$ (say, $L/c > 10$ times the resolution) and hence the minimum pulse duration ($\approx 5$ $L/c$). Here we choose $L/c = 80$ ns ($L \approx 24$ m) and pulse duration 500 ns. The repetition rate is chosen 500 kHz as a compromise between the time needed to record sufficient statistics and the time allowed by the setup's stability (see the SM Section). Photons propagate from the source to the stations through single mode optical fibers. The distance as measured through the fibers is constant and equal to 21 m from the source to each fiber polarizer. Fiber polarizers are robust, stable and easy to align, but their contrast (1:100) is poorer than the typical one of crystal cubes. For this reason, the highest value of $S_{CHSH}$ that can be achieved is $\approx 2.77$.

The observation angles are set mechanically. Therefore, there is time enough for the settings' information to propagate across the setup while the photons are in flight. We assume this information to be lost when the source of entangled states is turned off between pulses (see H3 below). In consequence, during the early part of the pulses (i.e., during a time $\approx L/c$ after the pulses' start), the projections of the polarization states of each photon in each station do occur as separate events in space time.

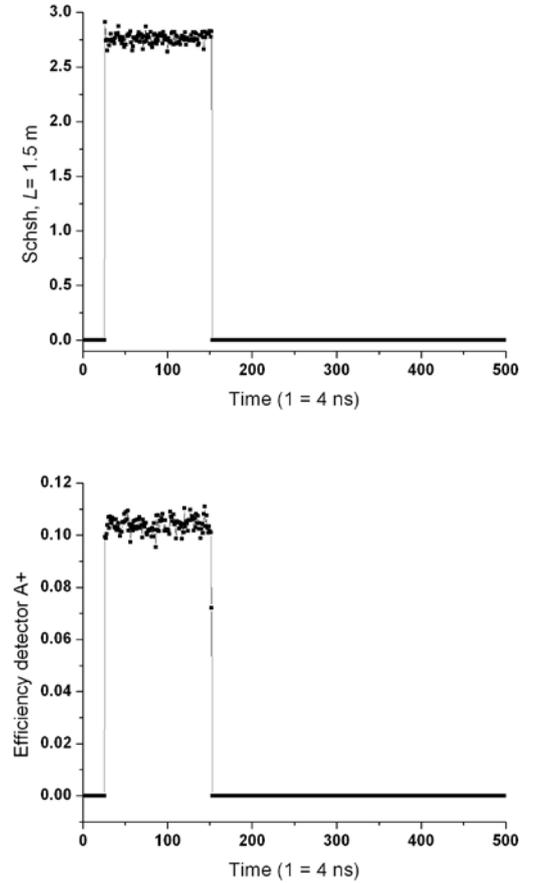

Figure 1: Up: $S_{CHSH}(t)$; down: $\eta(t)$ (detector A+), for distance between stations $L=1.5$ m in straight line, distance as measured through the optical fibers is 42 m. Pulse duration 500 ns, repetition rate 500 kHz. Full pumping period is displayed, time resolution is 4 ns.

As reference, time evolutions of entanglement and efficiency for short distance ($L= 1.5$ m, or $\tau \approx 5$ ns) are displayed in Figure 1. The efficiency is for the detector in the "+" output of the polarizer in station A. Efficiencies in the other detectors evolve in the same way. Time evolution of singles and coincidences follow the (roughly square) pulse pump shape; they do have some fluctuations that compensate when efficiency rates and correlations are calculated (see SM Section). That's why both $S_{CHSH}(t)$ and $\eta(t)$ have an almost perfect square shape with flat tops. For $L$=1.5 m, $\tau$ is close to the time resolution, so that no deviations are expected to be visible even if the TDH were true.

The (stroboscopic) time averaged value of $S_{CHSH}(t)$ is 2.762 with time dispersion 0.047; it agrees with the value calculated by taking into account all available data: 2.762±0.002. In the same way, $\langle\eta(t)\rangle_{A+}$ = 0.104 with time dispersion 0.004, while $\eta_{A+}$ calculated from all data is slightly smaller: 0.0960±3×10$^{-4}$. The difference is because the latter includes data both inside and outside the pulses, thus increasing the number of single detections.

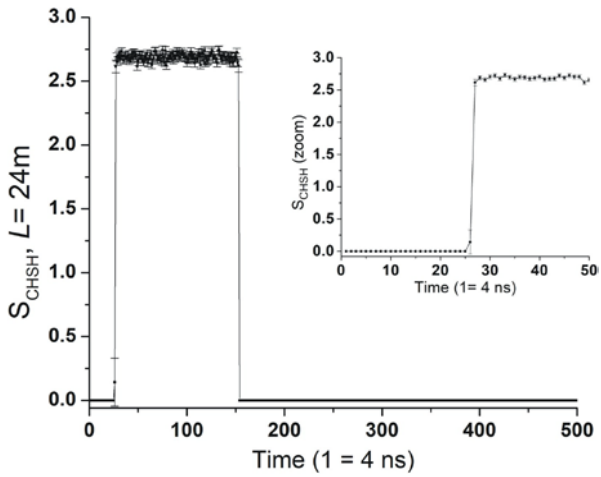

Figure 2: $S_{CHSH}(t)$ for $L$=24 m in straight line, full period. Inset: zoom on the first 100 ns of the pulse. Statistical error in each time slot is indicated. Other data as in Fig.1.

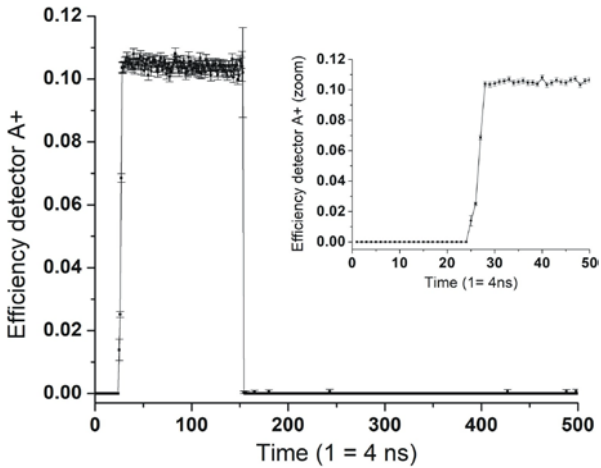

Figure 3: $\eta(t)$ (detector A+) for $L$=24 m in straight line, full period. Inset: zoom on the first 100 ns of the pulse. Statistical error in each time slot is indicated. Other data as in Fig.1.

The results obtained for $L$= 24 m ($L/c$≈ 80 ns), are displayed in Figures 2 and 3. Excepting for the different physical distance between the stations, everything is the same as for $L$=1.5 m. Statistical error for each time slot is plotted. No transient deviations from QM predictions in $S_{CHSH}(t)$ or $\eta(t)$ are observed. Both curves copy the rise of the pump pulse accurately (see the inset figures) and also its fall, in agreement with QM. According to the TDH, the shortest possible deviation should last 80 ns at least (or 20 "dots" in the figures) and should be hence clearly noticeable.

The time averaged value of $S_{CHSH}(t)$ is 2.688 with time dispersion 0.05; it is consistent with the value calculated using all data: 2.687±0.003. In the same way, $\langle\eta(t)\rangle_{A+}$ = 0.106 with time dispersion 0.045, while $\eta_{A+}$ calculated from all data is 0.0967±10$^{-4}$.

It is important defining the hypotheses involved in this experiment. The observations demonstrate that at least one of them is false.

H1) Transient Deviations: The bound in Eq.1 holds for times shorter than $L/c$.

H2) Stroboscopic: the physical system is in the same (or dynamically equivalent) "ground state" before the start of each pulse, and evolves in the same way when each pulse starts. In our setup, the system is assumed to decay to this "ground state" in a time <1.5 μs (≈19 $L/c$) after the end of each pulse (see SM Section).

H3) No correlation: when the source is turned off, the physical systems at the stations and the source are uncorrelated in spite there is enough time (at the speed of light) to interchange information among them (say, conspiratorially).

H4) Fair sampling: the measured value of efficiency is proportional to the value that would be observed with ideally efficient detectors. In other words: the detectors do not conspire to hide a transient evolution of $\eta(t)$, if it exists.

The "direct experiment" to test H1 does not require the stroboscopic assumption H2 but implies a major effort, of the order of a LIGO or a super collider. F.ex: supposing that the highest reported rate of coincidences in the lab [6] is somehow increased 10$^3$ times and, as it is done here, $\tau$ is divided in 20 slots; then $L \approx$ 20 km. The experiment described here is a rational first step before engaging in such an effort.

The loophole-free experiments have demonstrated that Bell's inequalities are violated even if the action of the conspiratorial mechanisms is made impossible. Therefore, these mechanisms do not exist or, at least, the role they play in the violation of Bell's inequalities is not relevant in those experiments. It seems improbable that in *this* experiment instead, they are as relevant as to make existing transients fully undetectable (by violating H3 or H4).

In summary: The requisites on time resolution, distance between stations and size of the statistical samples to detect the TDH are fulfilled, but we find no evidence of deviations from QM predictions. Strictly speaking, our findings disprove at least one of the hypotheses H1 to H4. Nevertheless, by reasonably extending the results of the loophole-free experiments, the space left to the validity of the TDH (i.e., H1) is reduced to the falsity of H2 (stroboscopic). Anyway, it would be desirable repeating the experiment with fast and unpredictable variation of the settings and high efficiency detectors (as cryogenic nanowires) in order to unquestionably discard the falsity of H3 and H4.

Experimental details and additional plots of the time evolution of singles, coincidences, $S_{CHSH}(t)$, $\eta(t)$ and $S_{CHSH}(t) \times \eta(t)$ are presented in the SM Section.

**Acknowledgments.**

This work received support from the grants N62909-18-1-2021 Office of Naval Research Global (USA), and PIP2022-0484CO and PUE 229-2018-0100018CO from CONICET (Argentina).

# SUPPLEMENTARY MATERIAL.

## 1. Description of the experimental setup.

It is sketched in Figure SM1. Biphotons at 810 nm in the fully entangled Bell state $|\varphi^+\rangle$ are produced in the standard configuration using two crossed (1mm long each) BBO-I crystals, pumped by a 40 mW diode laser at 405 nm with 40 mm coherence length (as measured at 100 kHz and 10% duty cycle; coherence length is observed to increase with duty cycle). This laser is able to emit fairly square pulses of adjustable duration and repetition rate, but shape deteriorates if the rate is higher than 1MHz or pulse duration is shorter than 200 ns. Duty cycles as low as 5% have been used with satisfactory laser's outcome. Low duty cycles are desirable for they give more time the system to decay to its "ground state" (see H2 in the main text) but, on the other hand, they also mean longer real time to gather sufficient statistics, which is a drawback because of birefringence compensation instability (see Sections 2 and 3 below).

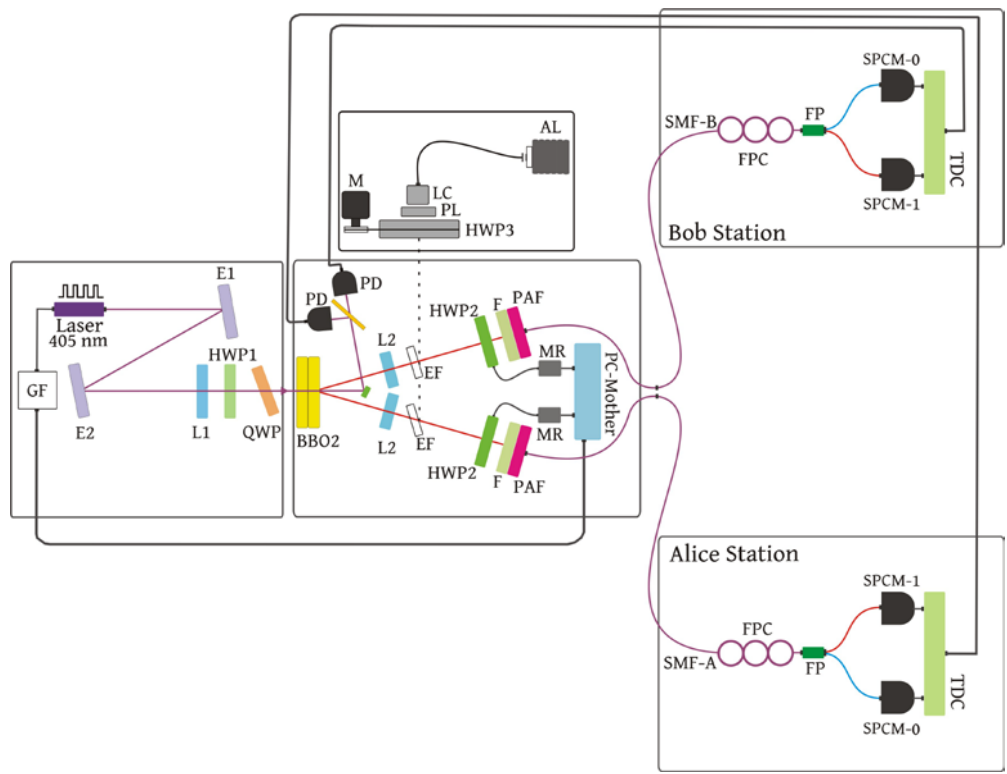

Figure SM1: Sketch of the setup. GF: function generator; L1,L2: f= 300 mm lenses; E1, E2: HR plane mirrors at 405 nm; HWP1 and QWP: half and quarter waveplates at 405 nm; BBO2: crossed BBO-I crystals (source of entangled states); PD: fast photodiodes, they send trigger signals to the TDCs via coaxial 50Ω cables; HWP2, HWP3: half-waveplates at 810 nm; F: Interferential filters at 810 nm, Δλ=10 nm, 90% transmission (according to specs); EF: removable HR plane mirrors in flip-flop mountings; AL: auxiliary CW laser diode at 810 nm coupled to multi-mode fiber; LC: collimating optics; PL: linear polarizer; M: motor that rotates HWP3; MR: servo motor controllers of HWP2; PAF: fiberports f = 7.5 mm; SMF-A and B: single-mode fiber coils, 21 m total length each; FPC: birefringence compensator ("bat-ears"); FP: fiber polarization analyzer; SPCM: photon counting module; TDC: time-to-digital converter (Id Quantique, Id-900).

A sample of the laser beam is sent to a 50-50 beam splitter. The output beams illuminate two fast photodiodes, which send electrical signals to each station indicating the start of each pulse. These signals propagate through 50Ω coaxial cables of 38 m length, and are checked to have negligible distortion. Two photodiodes are used (instead of just one and a coaxial "T") because spurious echoes in the long cables were observed otherwise. Trigger signals are stored in the #3 input channels of the time-to-digital converters (TDCs, Id Quantique Id-900). These are the largest files, because most pulses are "empty": only ≈2% of the pulses produce detected photons. This low number is necessary to keep the number of accidental coincidences low in the pulsed regime [1]. In a typical recording run, tens of millions of trigger signals must be recorded correctly by both TDCs, what is a difficult challenge. In order to keep tracking of pulse numbering with independent clocks (which unavoidably drift away), the repetition rate is switched or modulated, in order to establish a "physical" synchronization between the clocks in the TDCs [2]. The pulsed regime refreshes the synchronization between the distant clocks with each pulse; the frequency modulation allows reliable pulse numbering and immediately determines the correct delays between the lists of photons' detections without need of counting coincidences.

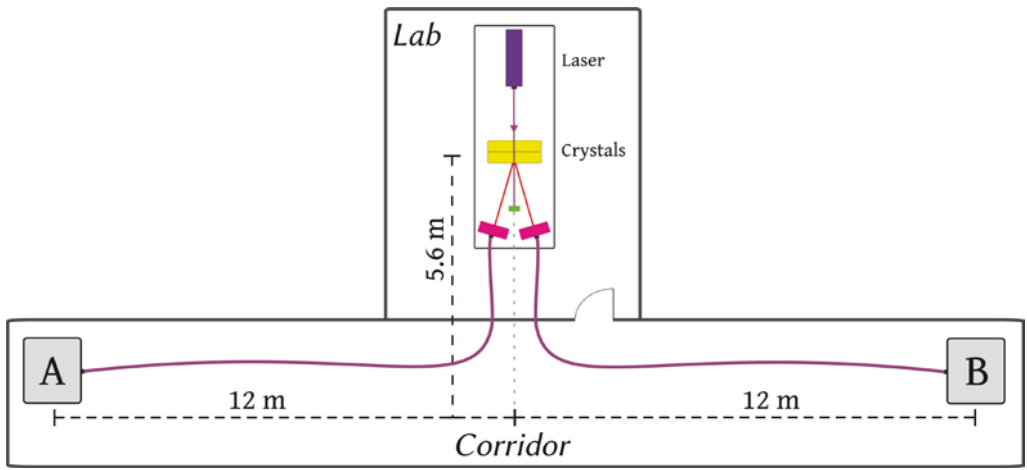

Figure SM2: Sketch of the experiment's positions in space. The pump laser, crystals and optics are on an optical table inside the Lab. Optical fibers are inserted in stainless steel tubes that provide mechanical protection and isolation from spurious light. The tubes have one end fixed to the optical table, pass through one of the Lab's walls and enter the adjacent corridor. The tubes extend to stations "A" and "B", where time tagged values of photons' detections and pulses' emission times are recorded using identical devices (the TDCs). Coaxial cables carrying information of the pulses' emission pass through the wall too, but are not indicated in this Figure. Distance from the crystals to the center of the corridor is 5.60 m. Distance between the point where the imaginary line of the pump laser intersects the center of the corridor, and each of the centers of the fiber polarizers in each station, is 12 m, then $L$= 24 m. In the case of $L$= 1.5 m the stations are moved inside the lab, and the tubes (which must be partially coiled) entered through the door.

    The entangled beams propagate through single-mode optical fibers (S630-HP Nufern) 21 m long each, which are extended from the source to the stations. The fibers are inserted into flexible stainless steel tubes (12 mm inner diameter, 16 mm external), which traverse the lab's walls through drilled holes to the adjacent corridor and reach the stations, see Figure SM2. All optics in each station is placed inside a box that protects from spurious light and dust. The stainless steel tubes end into these boxes. When measuring at short $L$, the stations are moved inside the lab and the tubes bent to re-enter the lab through the door.

    In each identically equipped station, "bat-ears" are used to compensate birefringence distortion in the fibers. Polarization is measured with two exit fiber optic analyzers (Thorlabs PFC-780SM-FC). The fibers are not a single piece, but three: from the focusing optics at the source 15 m, then a FC connector to a 5 m patch that includes the section coiled in the bat-ear, then another FC connector, and finally 1m of fiber up to the position where projection of the polarization state occurs. This amounts to 21 m of fiber for each station. Transmission in the fibers (from the output of the focusing fiberports optics until the detectors) is measured using the auxiliary laser: 83% for Alice and 82% for Bob.

    The beams leaving the two outputs of the fiber polarizers are sent to single photon counting modules (SPCM, AQR-13 and AQRH-13, from Perkin-Elmer-Pacer-Excelitas). These modules emit one TTL signal for each detected photon. The TTL signals are sent to channels #1 and #2 of the TDCs in each station. Detections' time values are stored. The TDCs have 10 ps nominal time resolution, but accuracy is reduced to ≈2 ns because of detectors' jitter. One PC in each station

controls the duration of the observation run (see below), the opening of files and their naming and saving, following the instructions sent by a "Mother" PC placed near the source.

"Mother" directly controls the function generator that pulses the laser (including the switching or modulation of the repetition rate, following a previously specified program) and the servo motors that adjust the settings angles. She also controls remotely wifi through a TCP/IP communication via a local network, the "sons" PCs in each station (Alice and Bob) to open, name and close the data files recorded in each TDC. Raw data are saved in .BIN format.

The coaxial cables carrying the pulses' signal are 38 m long each, but (as said) the fibers are only 21 m long each. This means that "trigger" signals arrive to the TDCs later than "signal" photons. Yet, this is not a problem, for the TDCs record data continuously in all the input channels. The delay is constant ($\approx$57 ns) and taken into account during data processing. Photons' detection times are positioned in reference to the trigger. Note that, in each station, time values in all channels are measured by a single clock. Synchronization between the clocks in each station is achieved through the trigger pulses arriving to channels #3, as explained.

**2. Recording data.**

One "run" is an interrupted session of recording data in the three channels in each station. Data are saved at the end of each run. Once the controlling program is started, the setup is able to perform an arbitrary number of successive runs with different settings (which are previously loaded in a .txt file in Mother) without the operators' assistance.

Each run records data during 30 s of real time. Recording runs are gathered in sets named "experiments", which accumulate the results of 32 runs (these are 8 for each of the 4 angle settings necessary to calculate $S_{CHSH}$, which are repeated cyclically). Because of dead periods introduced to give time the PCs in the stations to save and unload the data files reliably, each experiment lasts almost one hour.

Having well separated pulses is desirable to give the system a time as long as possible to decay to the "ground state" hypothesized in the stroboscopic approach. This means a low duty cycle and hence a short time of effective recording of data, what implies a long total real time of observation. As discussed in the main text, a minimum of $\approx 10^3$ coincidences must be recorded *in each time slot* to get sufficient statistics. But, if the duty cycle is made too small, the total real time of observation (necessary to record that amount of coincidences per slot) will become larger than the time birefringence compensation remains stable (see next Section 3). We find a compromise value of 25% duty cycle, what gives the systems in the stations 1.5 μs ($\approx$19 $L/c$) to decay to the ground state before the arrival of the next pulse (which may be "empty"). It would be desirable having a brighter source to repeat the experiment with a smaller duty cycle.

A measuring session starts by performing a checking experiment with 34 different setting angles, to get complete set of curves of total coincidences as functions of setting angles, see an example in Figure SM3. Before going on, these curves are rapidly analyzed to check that they are not shifted or distorted because of incomplete birefringence compensation, and an estimation of the (not time resolved) value of $S_{CHSH}$ is calculated.

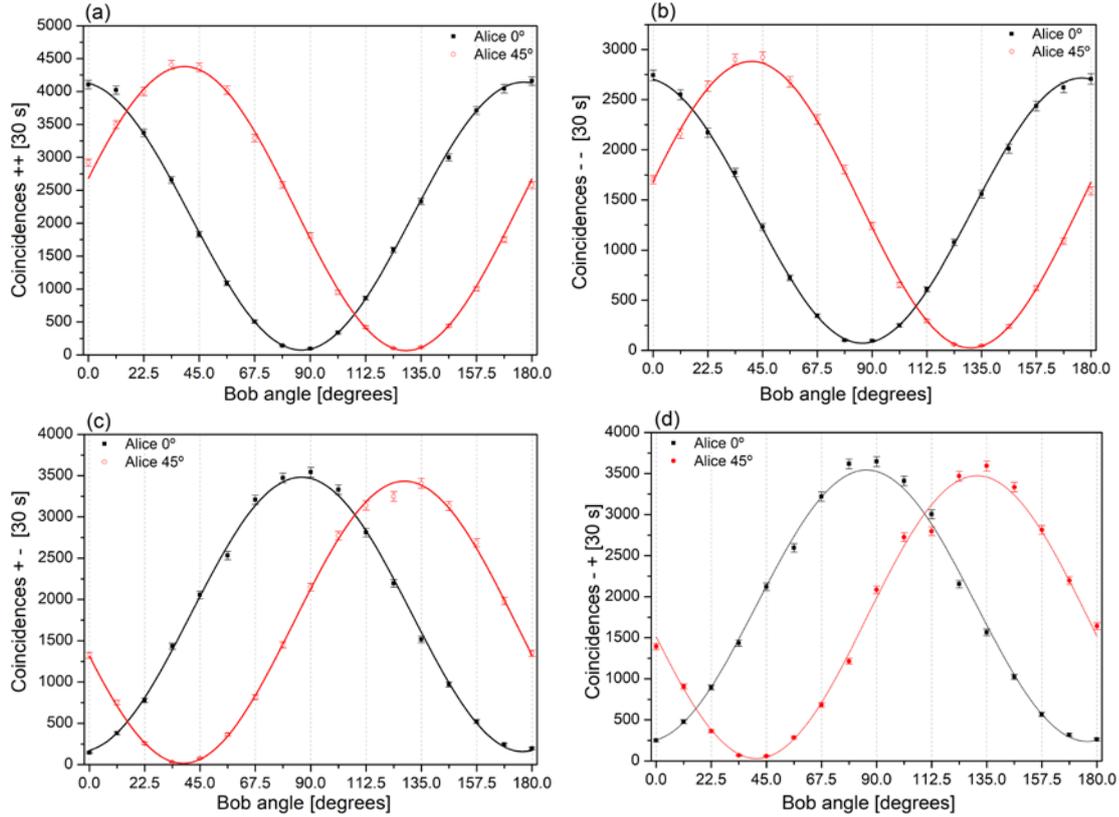

Figure SM3: Illustration of the curves recorded in a checking experiment (#181 in this case, *L*=24 m) to control that alignment and compensation birefringence is satisfactory; (a) Total number of coincidences as a function of the setting angles for "+,+" coincidences (i.e., coincidences between the detectors "+" in each station), (b) for "-,-", (c) for "+,-", (d) for "-,+". Measured $S_{CHSH}$ = 2.75, compare with $S_{CHSH}$= 2.69 measured in the experiments #182-188 that followed during the same day. The small 0.06 drop is caused by the variation of birefringence during the day, see Section 3.

If the results are not satisfactory, realignment and/or improved birefringence compensation (see next Section 3) are performed. The checking experiment with 34 settings is then repeated. If everything is satisfactory instead, several experiments using the 4 angle settings necessary to calculate $S_{CHSH}$ are carried out. Data of at least 4 experiments, using the 4 settings angles (each one repeated 8 times, cyclically), are accumulated to gather enough statistics. Sometimes a glitch in one of the PCs damages the data recording and/or saving processes, and part of the data must be discarded. In general, more than 4 experiments are necessary to gather sufficient statistics. If birefringence compensation is modified, the following experiments belong to a different session. We do not sum up data belonging to experiments recorded in different sessions.

**3. Birefringence compensation.**

In order to make easier the (largely try-and-error) method to compensate birefringence with bat-ears, we use an auxiliary laser diode at 810 nm. This is practically the wavelength of the entangled photons. The laser is fiber coupled (multi-mode). The beam is collimated, polarized and passed through a half wavelength waveplate. This waveplate is mounted in a motor rotating at ≈50 Hz. The result is a polarized beam at 810 nm which plane of polarization rotates at ≈100 Hz. This beam is inserted into the single mode fibers by using high reflection (at 45°) plane mirrors on flip-flop mountings (so that the mirrors can be easily removed from the optical axes, see Fig.SM1).

In each station, two photodiodes are placed at the exits of the fiber polarizer, and their outputs observed in an oscilloscope, see Figure SM4. The first coil in the bat ear (a quarter waveplate) is adjusted to maximize contrast of these sinusoids. This means the point in the Poincaré sphere is in the equator. Then the motor is stopped, and the position of the rotating waveplate is manually adjusted to the position that corresponds to one of the output fibers in the fiber polarizer (this position was previously marked on the mounting within which the waveplate rotates). Then the half waveplate in the bat ear is adjusted to maximize the signal (which is constant in time now) in one of the photodiodes, and to minimize the signal in the other. Then the waveplate is put into rotation again, and fine adjustments are introduced. The procedure is repeated at the other station.

Although this method is effective, thermal and mechanical perturbations affect birefringence during the day. In Fig.SM4, the decay of the value of $S_{CHSH}$ with real time is shown along the day (November $1^{st}$, 2023), during a period of about 10 hs. This puts a practical limit to the total duration of a measuring session.

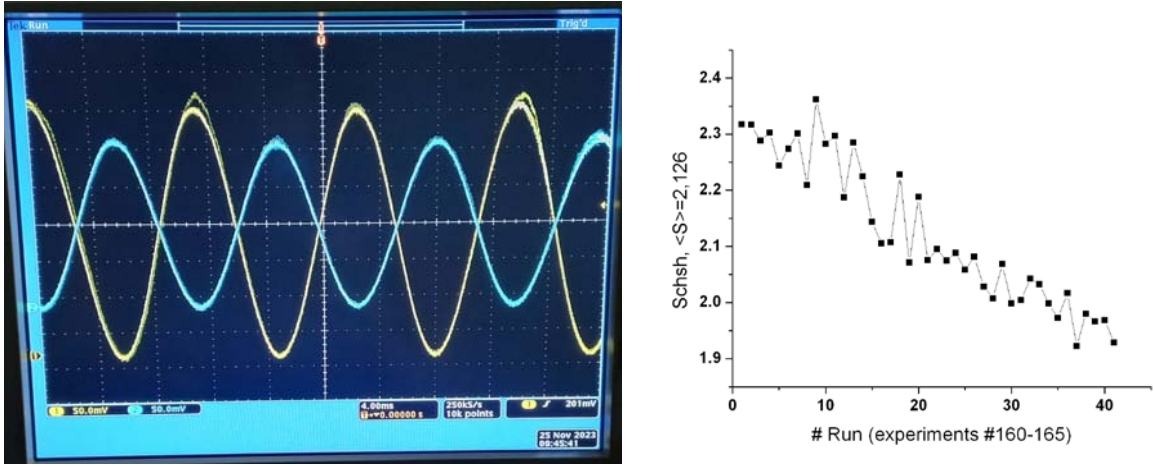

Figure SM4: Left: appearance of the photodiodes' signals when the auxiliary waveplate is rotating. Right: example of the decay of $S_{CHSH}$ along the day (about 10 hs) in a condition of deficient birefringence compensation.

Careful birefringence compensation using two photodiodes as described above is proven to be important to get values stable enough to record data during a whole day without having to introduce adjustments.

**4. Calculations, and some results.**

Photons' detections times are positioned in reference to the arrival of the trigger signal in channel #3. After summing up data produced by millions of pulses (typically $5\times10^8$ in a single experiment), plots of Singles and Coincidences as a function of time are obtained with sufficient statistics. Frequency modulation or switch of the pulses' frequency determines the numbering of each pumping pulse to be the same in each station, independently of clocks' drift [2]. Detections occurring during pulses with the same numbering are found to be coincident within 4 ns. There are practically no coincident detections observed outside the pump pulses, which agrees with the following estimation: for a 4 ns coincidence time window and detectors' typical dark count rate of 200 s$^{-1}$, the number of accidental coincidences accumulated during a 30 s run is: $(200 \text{ s}^{-1})^2 \times 4.10^{-9}$ s $\times$ 30 s $\approx 5\times10^{-3}$ in each time slot. Therefore, even if many runs are accumulated ($\approx$140 during one session), only rarely a slot outside the pulses has one coincidence.

As an illustration, stroboscopically reconstructed time variation of singles and coincidences for one of the detectors are displayed in Figure SM5. These curves match the laser pump pulse shape as observed with a fast photodiode.

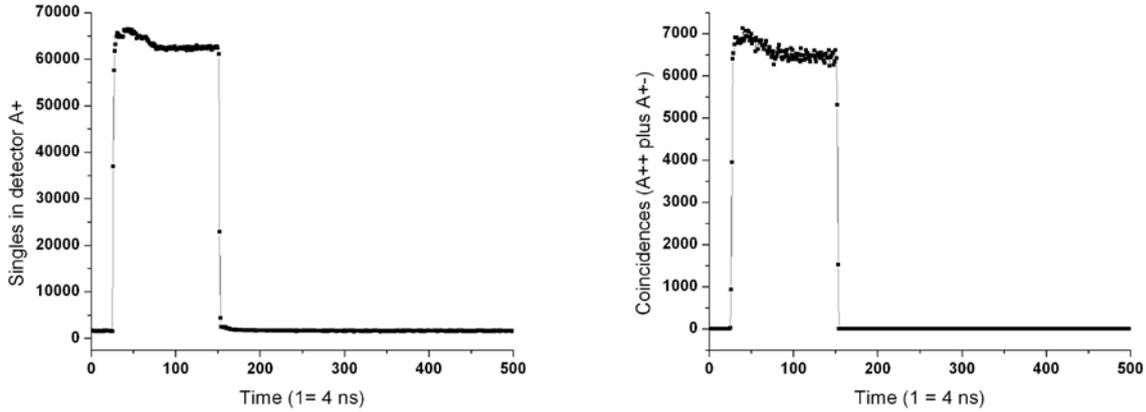

Figure SM5: Stroboscopic reconstruction of the time evolution of singles and coincidences in detector "+" in station Alice (A+). Left: single detections, Right: coincidences (A++ plus A+-). Note the number of coincidences in each time slot within the pump pulse is well above 1000, as required by the criterion of statistical significance described in the main text. The number of coincidences outside the pump pulses is zero, as expected. The number of singles outside the pump pulses ($\approx 1600$ /slot) is consistent with the rate of dark counts of detector A+ (140 s$^{-1}$).

The small fluctuations in the pulse shape observed in Fig.SM5 compensate when efficiency $\eta(t)$ and $S_{CHSH}(t)$ are calculated. Their curves are remarkably flat instead (see f.ex. Figure 1 in the main text). We have already observed this feature before [3-5].

The curves of the 16 types of coincidences recorded in experiments #182-188 ($L$=24 m) are shown in Figure SM6 below. They allow calculating $S_{CHSH}(t)$. It may be noted that for some types of coincidences and settings (say, +- for $\alpha=0$, $\beta=22.5°$), the number of coincidences per slot is smaller than the established criterion of statistical significance ($10^3$), yet the *total* number of coincidences (i.e., summing up ++,+-,-+,--), which is used to calculate $S_{CHSH}$, is always well above $10^3$ per slot.

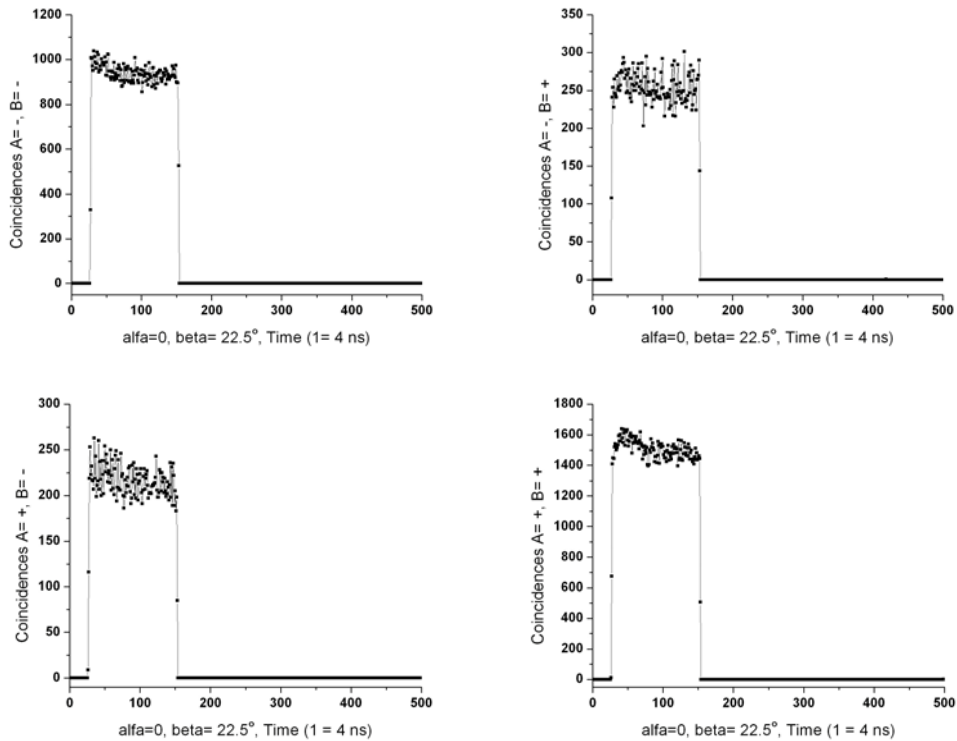

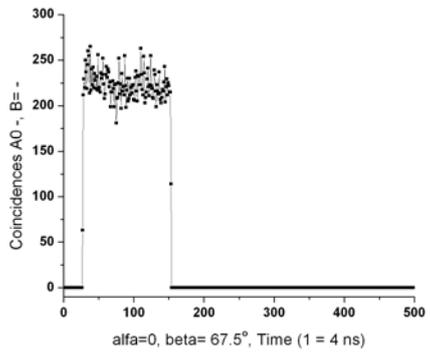
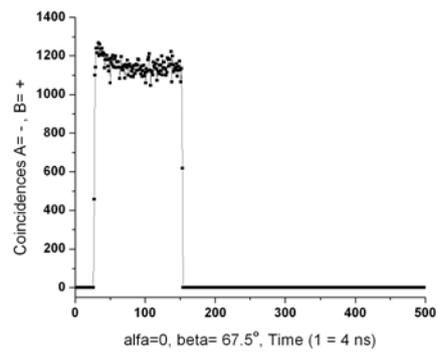
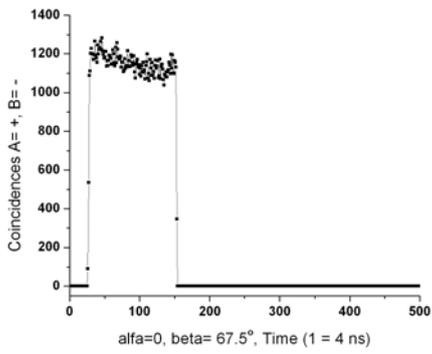
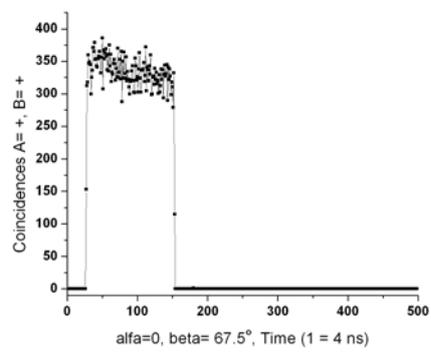
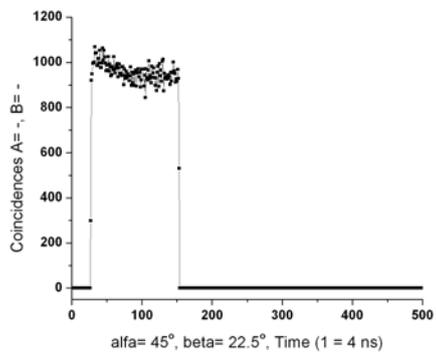
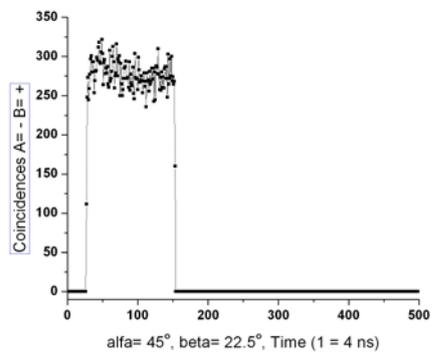
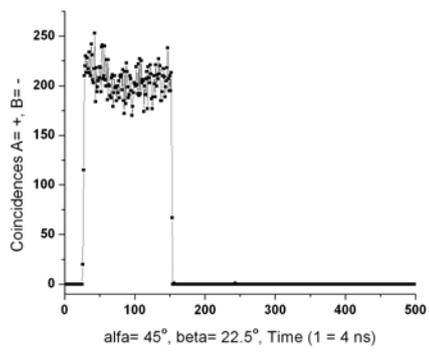
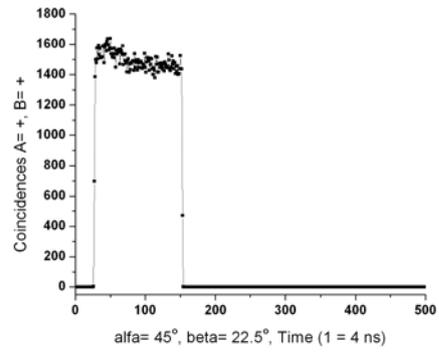

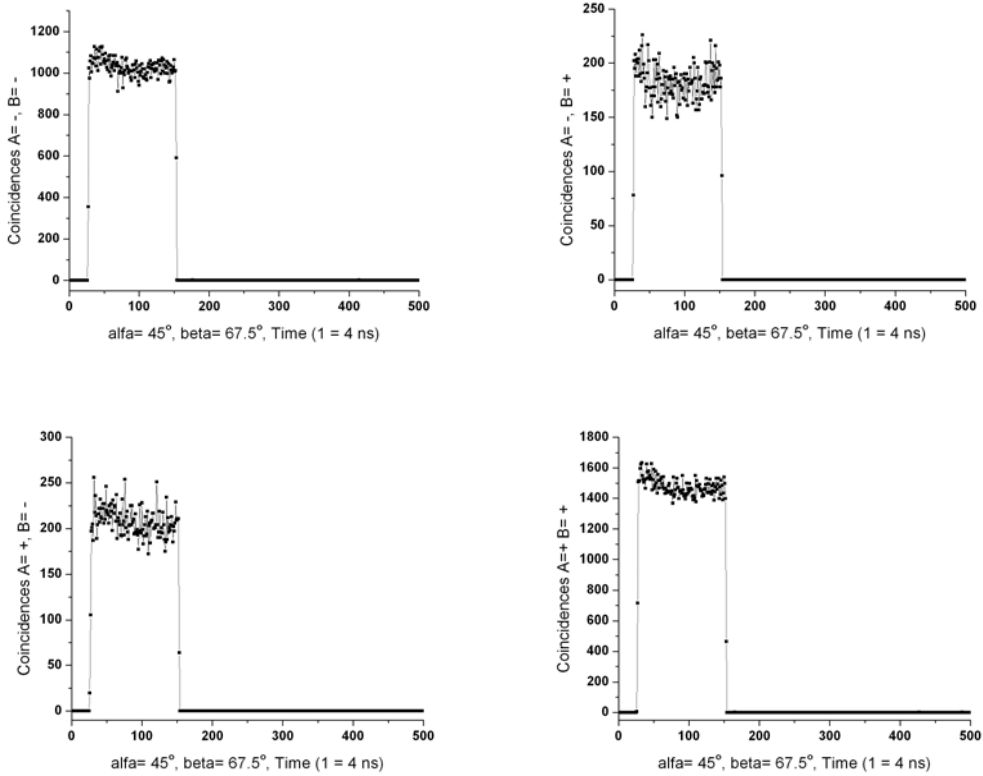

Figure SM6: Stroboscopic reconstruction of the time evolution of coincidences among all detectors for the case $L=24$ m. They allow the calculation of $S_{CHSH}(t)$ in Fig.2 in the main text. Full period of the pump laser is displayed.

For the sake of completeness, the curves $S_{CHSH}(t) \times \eta(t)$ (see Eq.1 in the main text) for both $L=1.5$ m and $L=24$ m cases are shown in Figure SM7. As can be anticipated from the figures in the main text, they are both flat squared curves with small statistical fluctuations, with no relevant difference between them. Because of the low efficiency of detectors and collecting optics $S_{CHSH}(t) \times \eta(t) < 2 \; \forall t$ in both cases, but we emphasize that the hypothesized transients, which should be easily visible in the $L = 24$ m case, are not observed. If H4 (fair sampling) is assumed valid, then the vertical axis can be rescaled with the measured value of efficiency $\approx 0.1$, and the tops of the figures become close to $2\sqrt{2}$ even for $t < L/c$, as it is predicted by QM.

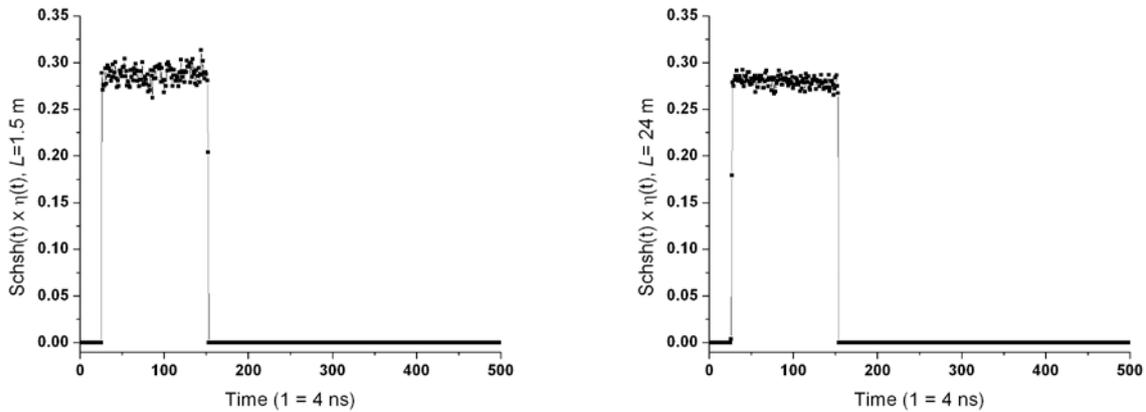

Figure SM7: Stroboscopic reconstruction of the time evolution of $S_{CHSH}(t) \times \eta(t)$, efficiency is calculated for detector A+; Left: $L=1.5$ m, Right: $L = 24$ m. Note that $S_{CHSH}(t) \times \eta(t) < 2 \; \forall t$ because of detectors' and optics' limited efficiency, but no difference between the curves, or indication of transients, are observed.

**References on experimental details (valid for this SM section only).**